\documentclass[11pt,twoside]{article}

\usepackage{asp2006}
\usepackage{epsf}
\usepackage{lscape}

\begin{document}
\title{Cosmic Infrared Background ExpeRiment (CIBER): A Probe of 
Extragalactic Background Light from Reionization}   

\author{Asantha Cooray$^1$, Jamie Bock$^2$, Mitsunobu Kawada$^3$,Brian Keating$^4$, Andrew Lange$^2$,
Dae-Hee Lee$^5$, Louis Levenson$^2$, Toshio Matsumoto$^6$, Shuji Matsuura$^6$, 
Tom Renbarger$^4$, Ian Sullivan$^2$, Kohji Tsumura$^6$, Takehiko Wada$^6$, Michael Zemcov$^2$}

\affil{
$^1$ Center for Cosmology, University of California, Irvine, USA\\
$^2$ Department of Physics, Caltech, Pasadena USA\\
$^3$ Department of Physics, Nagoya University, Japan\\
$^4$ Department of Physics, University of California, La Jolla, USA\\
$^5$ Korea Astronomy and Space Science Institute, Daejeon, Korea\\
$^6$ Institute of Space and Astronautical Sciences, JAXA, Japan\\
}    

\begin{abstract} 

The Cosmic Infrared Background ExpeRiment (CIBER) is a rocket-borne 
absolute photometry imaging and spectroscopy experiment optimized 
to detect signatures of first-light galaxies present during reionization in the unresolved 
IR background. CIBER-I consists of a wide-field two-color camera for fluctuation measurements,
 a low-resolution absolute spectrometer for absolute EBL measurements, and a narrow-band
imaging spectrometer to measure and correct scattered emission from the foreground zodiacal cloud.
CIBER-I was successfully flown on February 25th, 2009 and has one more planned flight in early 2010.
We propose, after several additional flights of CIBER-I, an improved CIBER-II camera consisting of a
wide-field 30 cm imager operating in 4 bands between 0.5 and 2.1 microns.  It is designed for a high significance detection  of unresolved IR 
background fluctuations at the minimum level necessary for reionization.
With a FOV 50 to 2000 times largerthan existing IR instruments on satellites, 
CIBER-II will carry out the definitive study to establish the surface density 
of sources responsible for reionization.

\end{abstract}



\section{Introduction}   

The optical and UV radiation from sources present during reionization
 is now present in the near-infrared with a small, but non-negligible, contribution to the
Extragalactic Background Light (EBL). 
Searches for this radiation based on absolute photometry have proven problematic due to confusion with the 
Zodiacal foreground.  
Instead of the absolute background, in Cooray et al. (2004), 
we proposed to develop a near-infrared sounding rocket experiment, CIBER, to conduct 
a deep search for extragalactic background fluctuations from the epoch of reionization associated with
first-light galaxies. 

\begin{figure}[!ht]
\begin{center}
\resizebox{0.8\hsize}{!}{
    \includegraphics{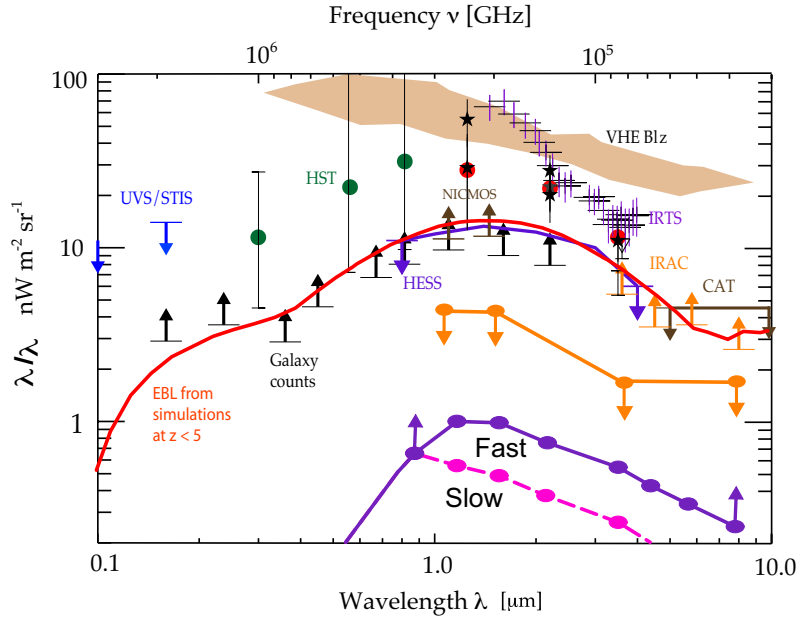}
}
\end{center}
\caption{
Summary of EBL observations at infrared and optical wavelengths, showing upper limits, reported residuals after subtraction of local foregrounds, reported detections with absolute photometry (DIRBE and IRTS), and integrated galaxy counts (lower limits). The large experimental scatter at 1.2 $\mu$m is notable, with Zodiacal removal suspected as a prime source of systematic error (Dwek et al. 2005).  A recent upper limit based on TeV absorption using HESS (Aharonian et al. 2006) contradicts the excess reported by several authors at 1-2 $\mu$m. The indirect EBL measurements with TeV spectra, however, do not provide consistent results since a previous independent estimate suggests a higher background (the region labeled VHE Blz; Schroedter 2005) . The red line shows the integrated galaxy light associated with galaxies formed at $z < 5$ based on a semi-analytical model (Primack et al. 2008). The purple and pink lower lines show the estimated EBL from $z > 6$ for fast and slow reionization histories, lower limits since the calculation is based on the minimum UV luminosity density needed to reionize and maintain the ionized state of the intergalactic medium (Chary \& Cooray 2009). The orange upper limits show constraints on the first-light EBL from reported fluctuations in deep Spitzer and NICMOS images.
}\label{fig:satellite}
\end{figure}

The EBL spectrum contains all radiative information from the reionization epoch (Santos et al. 2003; 
Kashlinsky et al. 2004; Fernandez \& Komatsu 2006; Cooray et al. 2009). 
We expect that the EBL 
contains diffuse signatures of reionization, such as Ly-$\alpha$ background radiation redshifted to 
near-IR wavelengths today. Remnants of these first stars, black holes in the form of miniquasars  
will also contribute to the EBL (Cooray \& Yoshida 2004).

Interestingly, integrated individual galaxy counts appear to fall short of the EBL measured with absolute photometry
 at near-infrared wavelengths (Fig.~1). For instance in the 1-3 $\mu$m 
band, galaxies contribute an intensity 
of $\sim$10 nW/m$^2$ sr (Madau \& Pozzetti 2000). 
In contrast, the extragalactic background light (EBL) measured by DIRBE and the IRTS 
at the same wavelengths ranges from 10 nW/m$^2$ sr at 3.6 $\mu$m (Levenson \& Wright 2008)
to up to 60 nW/m$^2$sr at 1.2 $\mu$m (Cambresy et al. 2001; Matsumoto et al. 2005).
It is highly unlikely that this whole difference is associated with sources during reionization, since such
a large intensity requires unphysical requirements on starformation (Madau \& Silk 2005).

Though the total luminosity produced by sources responsible for reionization is uncertain, 
a lower limit can be placed assuming a minimal number of photons to produce and maintain reionization,
given existing information on reionization, rest-frame UV luminosity functions of galaxies at
$z > 6$, stellar mass estimates for galaxies at $z\sim 6$, among others (Chary \& Cooray 2009).
Such ‘minimal’ reionization scenarios produce an EBL $\sim$ 1 nW/m$^2$ sr, a level undetectable by 
current absolute photometry measurements which, after dedicated space-borne 
measurements, show large discrepancies. 
As the spectral signature in EBL from reionization  contains integrated emission from all sources, including fainter 
ones undetectable with JWST, it captures the exact reionization history and provides more
information than other known probes of reionization, including CMB and 21-cm background.
This spectral feature could be resolved in the future with absolute photometric measurements in narrow spectral 
bands between 1-2 $\mu$m with an out-of-Zodi EBL explorer at distances around 5 AU (Cooray et al. 2009).

\begin{figure}[!ht]
\begin{center}
\resizebox{0.9\hsize}{!}{
    \includegraphics{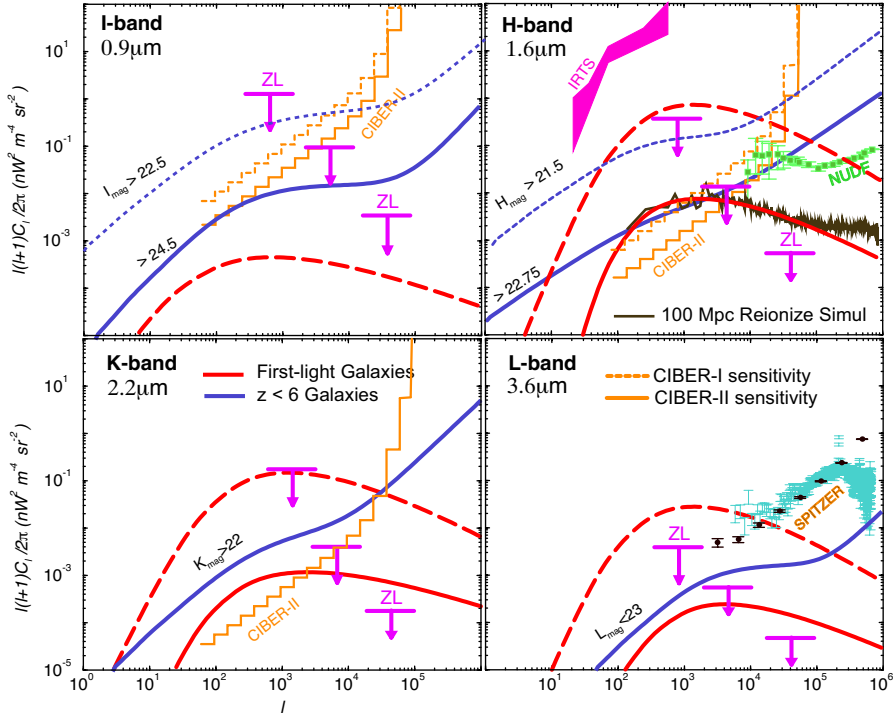}
}
\end{center}
\caption{
Spatial power spectrum of EBL fluctuations in standard IR bands. The red curves show the power spectra of fluctuations from first-light galaxies forming over the redshift interval $8 < z < 15$.  The top red dashed line shows the case where fluctuations are normalized to {\it Spitzer} measurements (Kashlinsky et al. 2008).
The bottom solid line shows the minimal signal necessary to produce reionization.
Cyan points show {\it Spitzer} measurements by Kashlinsky et al. while black points show residuals from Cooray et al. 
after subtracting faint, blue dwarfs detected in deep HST ACS images.  The dark blue curves give estimated 
fluctuations from known galaxies, as a function of magnitude cutoff, based on a galaxy distribution model 
based on the halo appoach (Cooray \& Sheth 2002; Cooray 2006) matched 
to existing clustering data (e.g., Sullivan et al. 2007).  
The galaxy cutoff taken for CIBER is 25\% pixel 
removal using deep ancillary source catalogs.  CIBER-II (solid blue) has a lower residual local galaxy foreground due to its smaller pixel size compared with CIBER-I (dashed blue).  The orange stair-steps in three panels shows the 
binned statistical sensitivity of CIBER-I and CIBER-II instruments in a single 50s observation.  
Zodiacal light is known to be spatially uniform, shown by the upper limits.
The black line shows the EBL fluctuations resulting from a 2048$^3$ particle, large volume (100 Mpc) numerical simulation of reionization completed by the Princeton group (Trac \& Cen 2007) with a combination of both Pop II and Pop III stars in first-light galaxies, as analyzed by the UCI team.   The shape of fluctuations is model independent and shows the overall bump at $\ell \sim 1000$ in agreement with the analytical model.
}\label{fig:satellite2}
\end{figure}

Even in the relatively recent times since reionization it appears there may be an under accounting of galaxies.  
The current star-formation in Lyman-break galaxies (LBGs) at z $\sim$ 6  (Bouwens et al. 2006) 
falls about a factor of 6 to 9 below the 
minimum required to maintain an ionized IGM given canonical estimates for the clumping factor of the 
gas and the escape fraction of ionizing photons from galaxies. 
The upper limits on the ultraviolet luminosity 
function suggests a negative evolution with increasing redshift, in the sense that galaxies at the bright end 
of the UV luminosity function at z $\sim$ 6 were fainter in the ultraviolet. 
 The implication is that the contribution from star-formation in faint galaxies, below the detection 
threshold of current surveys, is higher than has been estimated. The {\it Spitzer} stellar mass estimates
suggest a to-heavy IMF for high redshift galaxies (Chary 2008), 
suggesting a different stellar population in those galaxies
than in the local Universe. If current galaxy surveys under account for galaxy evolution or miss populations, the 
deficit may be uncovered in a careful study of the EBL.

A first-light galaxy EBL component from reionization and any contribution from faint sources at $z \sim 6$ unresolved
by existing deep, pencil surveys can be uncovered using a careful study of background fluctuations (Cooray et al. 2004).  
The technique involves the measurement of the angular power spectrum via an anisotropy study, 
similar in method to well-established techniques for studying CMB anisotropy.  
First-light galaxies have predictable clustering, determined by the growth of dark matter perturbations in 
$\Lambda$CDM cosmology, and will produce EBL fluctuations with a characteristic spatial power spectrum.  
Zodiacal light is known to be spatially uniform on arcminute scales.  
What is uncertain is the amplitude of fluctuations since it depends on the astrophysics of reionization
and the importance of, say, fainter sources at $z \sim 6$ to 8 that are responsible for maintaining
reionization. A fluctuation measurement has the potential to detect a first-light galaxy component to the EBL at 
much fainter  levels than can be probed with absolute photometry.

Since our work on CIBER began (Cooray et al. 2004; Bock et al. 2006), 
{\it Spitzer} IRAC has carried out a study of EBL clustering anisotropy.  
After a deep removal of both point sources, and deconvolving emission from the extended PSF, 
Kashlinsky et al. (2005; 2007) claim detection of first-light galaxy fluctuations based on a deviation 
from a shot-noise power spectrum on the largest angular scales. 

Some doubt may be cast on the interpretation of first sources due to the angular scales and wavelengths 
covered by IRAC. Due to IRAC’s small field of view, the power spectrum from high-z objects must be 
carefully separated from fainter, unresolved foreground galaxies (Cooray et al. 2007; Chary et al. 2008).
In addition,  fluctuation measurements have been made at shorter near-IR wavelengths in narrow, but very deep, 
NICMOS images (Thompson et al. 2007). The images again show a fluctuation signal (Fig~2).  The 
spectrum of fluctuations measured between {\it Spitzer} IRAC and HST NICMOS appears to be approximately flat, 
and the authors state it does not correspond to the shape expected for $z > 8$ sources (Thompson et al. 2007), 
but rather to a  low redshift population unresolved in both {\it Spitzer} and HST.  

Considering the difficulty of the measurement, 
we believe the amplitude between NICMOS and {\it Spitzer} is quite uncertain.  We also 
note that narrow fields such as GOODS and NICMOS UDF are affected by cosmic variance, especially at $z > 6$, where 
the correlation length of halos responsible for reionization is greater than the field size.
These caveats do not rule out the possibility that the {\it Spitzer} fluctuations, at least
at some fractional level, reveal a first-light component.  
The initial results are clearly exciting, have important implications on galaxy formation, but 
IR background fluctuations must be further studied at shorter wavelengths, where 
the signatures from reionization are expected to dominate.

\section{CIBER-I} \label{sec:satellite}

The CIBER instrument currently consists of two wide-field imagers to measure EBL fluctuations, 
a Low-Resolution Spectrometer (LRS) to probe the EBL via absolute photometry, and a Narrow-Band Spectrometer (NBS), 
to measure the brightness of the Zodiacal by the Ca-II 854.2 Fraunhofer line (Bock et al. 2006).
Two imaging cameras operate at 0.9 and 1.6 $\mu$m and probe EBL fluctuations over 
a wide 4 sqr. degree field of view, allowing measurements over the 
distinctive peak in the power spectrum at $\ell \sim 2000$ (0.1 degrees).

With the NBS, CIBER will also test the large EBL intensity ($\sim$ 50 nW/m$^2$ sr) reported by DIRBE and IRTS.  
The measurement is not limited by the sensitivity of the instruments, but primarily by the Zodiacal foreground and
to a less extent by the astrophysical systematic errors in  
removing stars and scattered starlight, expected to be $\sim$ 2\% of the Zodiacal brightness.

As the residual EBL spectrum closely resembles that of Zodiacal light (Dwek et al. 2005), there is the
possibility of a large $\sim$ 25\% error in the Zodi model used by both DIRBE and IRTS teams.
The NBS has sufficient sensitivity and systematic error control to precisely measure the Zodiacal amplitude at 854.2 
nm.  This measurement can be extrapolated to the DIRBE 1.2 and 2.2 $\mu$m 
bands based on the Zodiacal spectrum measured 
by the LRS.  CIBER-I has a second planned flight in early 2010. We also 
propose several additional flights to observe multiple lines of sight through the Zodiacal cloud, measured between a span of 
6 months, to vary the solar elongation angle.  The observation windows 
are carefully chosen so these fields are within the DIRBE observations at the same time of year.  If the 
DIRBE models are incorrect at the 20\% level, this should become readily apparent, and the multiple observations 
are necessary to  help us understand how the Zodi models should be corrected.

CIBER-I was first launched from White Sands Missile Range in New Mexico on a Terrier-Black Brant sounding rocket on 
February 25th, 2009.  The vehicle performed as expected, and the payload achieved an apogee of 320 km.  All 
flight events went according to plan and we hope to release our first science results within a year.
These are expected to be on a first measurement of the EBL between 1-2$\mu$m with CIBER and on a fluctuation
analysis to test the claims of Kashlinsky et al. (2007) at shorter near-IR wavelengths.

\section{CIBER-II} \label{sec:satellite2}

After the planned flight 2010 and subsequent proposed flights with CIBER-I are completed, we plan a more capable camera designed to probe fluctuations down to the low level of minimal reionization fluctuations, with a factor of $\sim 10$ improvement in sensitivity compared to that of CIBER-I (see Fig.~2).  The CIBER-II camera consists of a 30 cm telescope operating simultaneously in four bands between 0.5 and 2.1 $\mu$m with each imaging the same 2 sqr. degree field of view.  
The four bands are matched to
identify the spectral dependence of the reionization contribution from foreground and Zodiacal fluctuations.
As the case with CIBER-I, CIBER-II will image wide fields with existing deep coverage in optical and near-IR with
{\it Spitzer} and ground-based instruments.

The cameras are designed for high sensitivity to surface brightness in the short amount of time available during a sounding rocket flight. In comparison with space-borne telescopes, CIBER-II measures 
fluctuations on large angular scales, on both sides of the expected peak in the power spectrum.  These are also the angular scales best suited for distinguishing this signal from local galaxies, and systematic effects due to its distinctive power spectrum.  By measuring large angular scales, potential problems with source removal are less serious than in a small deep measurement.  CIBER-II is designed to have a very large instrumental FOV, the figure of merit relevant for measuring surface brightness, with required
 raw sensitivity even in a short sounding rocket flight compared with the long integrations possible on a satellite. 
It is expected that CIBER-II will carry out the definitive 
study to establish the surface density of sources responsible for reionization
and its results will complement number counts at $z > 6$ from JWST, especially at the bright end,
to put a complete picture of reionization together.

\acknowledgements 
CIBER-I is funded by NASA APRA NNG05WC18G (at Caltech) and NNX07AG43G (at UCI). AC thanks funding from NSF CAREER 
AST-0645427, Award 1310310 from Spitzer, and HST-AR-11241/11242 from STScI.


\end{document}